\documentstyle[bo99,epsfig]{article}

\title{Relativistic flows in blazars}

\author{Gabriele Ghisellini}                                                       
\affil{Osservatorio Astronomico di Brera, Via Bianchi 46, 
I--23807 Merate, Italy}
\begin{document}

\maketitle

\begin{abstract}
The radiation we observe from blazars is most likely the
product of the transformation of bulk kinetic energy into random energy.
This process must have a relatively small efficiency (e.g. 10\%) if jets
are to power the extended radio--structures.
Recent results suggest that the average power reaching the extended
radio regions and lobes is of the same order of that produced
by accretion and illuminating the emission line clouds.
Most of the radiative power is produced in a well localized region of the jet,
and, at least during flares, is mainly emitted in the $\gamma$--ray band.
A possible scenario qualitatively accounting for these facts
is the internal shock model, in which the central engine produces
a relativistic plasma flow in an intermittent way.
\keywords{Jets, AGNs, blazars, radiation processes:
synchrotron, inverse Compton, electron--positron pairs}               
\end{abstract}

\section{Introduction}
We believe that the continuum radiation we see from blazars
comes from the transformation of bulk kinetic energy, and possibly
Poynting flux, into random energy of particles, which quickly produce 
beamed emission through the synchrotron and the inverse Compton process.
This is analogous to what we believe is happening in gamma--ray bursts,
although the bulk Lorentz factor of their flow is initially larger.

Evidences for bulk motion in blazars with Lorentz factors between 5 and 20
have been accumulated along the years, especially
through the monitoring of superluminally moving blobs on the VLBI scale
(Vermeulen \& Cohen 1994),
and, more recently, through the detection of very large variable powers
emitted above 100 MeV (see the third EGRET catalogue,
Hartman et al., 1999), which require beaming for the source to
be transparent to photon--photon absorption
(e.g. Dondi \& Ghisellini, 1995).

The explanation of intraday variations of the radio flux, leading
to brightness temperatures in excess of $T_{\rm B}=10^{18}$ K 
(much exceeding the Compton limit) are instead still controversial
(Wagner \& Witzel 1995).
Interstellar scintillation is surely involved, but it can work only if the
angular diameter of the variable sources is so small to nevertheless
lead to $T_{\rm B}=10^{15}$ K, which requires either a coherent process
to be at work (e.g. Benford \& Lesch 1998)
or a Doppler factor of the order of a thousand.

Another controversial issue is the matter content of jets.
We still do not know if they are dominated by electron--positron pairs
or by normal electron--proton plasma
(see the reviews by Celotti, 1997, 1998).

Part of our ignorance comes from the difficulty of estimating
intrinsic quantities, such as the magnetic field and the particle 
densities, using the observed flux, which is strongly modified 
by the effects of relativistic aberration, time contraction and blueshift, 
all dependent on the unknown plasma bulk velocity and viewing angle.
Furthermore it is now clear (especially thanks to multiwavelength campaigns)
that the blazar phenomenon is complex.

On the optimistic side, we have for the first time a complete information
of the blazar energy output, after the discovery of their $\gamma$--ray 
emission, and some hints on the acceleration process, through the behaviour 
of flux variability detected simultaneously in different bands
(see the review by Ulrich, Maraschi \& Urry 1996).
Also, blazar research can now take advantage of the explosion of 
studies regarding gamma--ray bursts, which face the same problem of how
to transform ordered to random energy to produce beamed radiation
(for reviews: Piran 1999; Meszaros 1999).

\section{Accretion = Rotation?}

% Extragalactic jets, as well as their galactic superluminal counterparts,
% may well carry more power than what is emitted by their accretion disk.
Despite the prediction that jets carry plasma in relativistic motion
dates back to 1966 (Rees, 1966), and intense studies over the last
20 years (Begelman, Blandford \& Rees, 1984),
quantitative estimates of the amount of power transported in jets have 
been done only relatively recently, following new observational results.

One important point is that the extended (or lobe) radio emission of 
radiogalaxies and quasars traces the energy content of the emitting region.
Through minimum energy arguments and estimates of the lobe lifetime by
spectral aging of the observed synchrotron emission and/or by
dynamical arguments, Rawlings \& Saunders (1991) found
a nice correlation between the average power that must be supplied to
the lobes and the power emitted by the narrow line region.
Although one always expects some correlation between powers (they both scales
with the square of the luminosity distance) it is the ratio of the two 
quantities to be interesting, being of order of 100.
Since we also know that, on average, the total luminosity in narrow lines
is of the order of one per cent of the ionizing luminosity, we have the
remarkable indication that the power carried by the jet (supplying the extended
regions of the radio--source) and the power produced by the accretion
disk (illuminating the narrow line clouds) are of the same order.

Celotti, Padovani and Ghisellini (1997) later confirmed this
by calculating the kinetic power of the jet at the VLBI scale
(see Celotti \& Fabian 1993) and the broad line luminosity 
(assumed to reprocess $\sim 10\%$ of the ionizing luminosity).

A possible explanation involves the magnetic field being responsible 
for both the extraction of spin energy of a rotating black 
hole and the extraction of gravitational energy of the accreting matter.
Assume in fact that the main mechanism to power the jet is the 
Blandford--Znajek (1977) process:
\begin{equation}
L_{\rm jet} \, \simeq \, \left( {a\over m}\right)^2 U_{\rm B} (3R_{\rm s})^2c
\end{equation}
where $(a/m)$ is the specific black hole angular momentum ($\sim 1$ for a
maximally rotating Kerr hole), $U_{\rm B}$ is the magnetic energy density
and $R_{\rm s}$ is the Schwarzchild radius.
Note that Eq. 1 has the form of a Poynting flux.
Assume now that most of the luminosity of the accretion disk is produced
at $3 R_{\rm s}$. 
The corresponding radiation energy density is then 
$U_{\rm r} = L_{\rm disk} / (36 \pi R_{\rm s}^2 c)$, leading to
\begin{equation}
L_{\rm disk} \, =\, U_{\rm r} (3R_{\rm s})^2c
\end{equation}
Therefore a magnetic field in equipartion with the
radiation energy density of the disk would lead to
$L_{\rm jet} \sim L_{\rm disk}$.

\section{Mass outflowing rate}

We can estimate the ratio of the outflowing (in the jet) to the 
inflowing mass rate, since
\begin{equation}
L_{\rm disk} \, =\, \eta \dot M_{\rm in} c^2; \,\,\,
L_{\rm jet} \, =\, \Gamma \dot M_{\rm  out} c^2;\qquad  \to \qquad
\dot M_{\rm  out}\, =\, { \eta \over \Gamma } 
{L_{\rm disk} \over L_{\rm jet} }  \dot M_{\rm in}
\end{equation}
If jets carry as much energy as the one produced by the accretion disk, 
we then obtain that the mass outflow rate is $\sim 1\%$ of the accreting 
mass rate (if $\eta = 10\%$ and $\Gamma=10$).

\section{The blazar diversity}

BL Lac objects and Flat Spectrum Radio Quasars (FSRQ) are characterized 
by very rapid and large amplitude variability, power law spectra in 
restricted energy bands and strong $\gamma$--ray emission.
These common properties justify their belonging to the same blazar class.
However they differ in many other respects,
such as the presence (in FSRQ) or absence (in BL Lacs) of broad emission lines,
the radio to optical flux ratio, the relative importance of the $\gamma$--ray
emission, the polarization degree, and the variability behavior.
Within the BL Lac class, Giommi \& Padovani (1994) have subdivided the objects 
according to where (i.e. at what frequency) the first broad
(synchrotron) peak is located.
Low energy peaked BL Lacs (LBL) show a peak in the IR--optical bands,
while in High energy peaked BL Lacs (HBL) this is in the X--ray band
(see, in this volume, the contributions
of Costamante et al., Giommi et al., Pian et al.,Tagliaferri et al.,
Tavecchio \& Maraschi, Wolter et al.).

As the emission of all blazars is beamed towards us, so there
must be a parent population of objects pointing in other directions.
The parent populations of BL Lacs and FSRQs are believed to be 
FR I and more powerful FR II radio galaxies, respectively (see
the review by Urry \& Padovani 1995).
The absence of broad emission lines in BL Lacs is shared by
FR I radio galaxies, whose nuclei are well visible
by Hubble Space Telescope observations
(Chiaberge, Capetti \& Celotti 1999).
This suggests that in FR I and BL Lac objects broad emission lines
are intrinsically weaker than in more powerful objects.

% \begin{figure}
% \vskip -2. true cm
% \centerline{\psfig{file=../../gro/2005f.ps, width=15cm}}
% \vskip -6 true cm
% \caption[]{The SED of the HBL PKS 2005--489 observed by {\it Beppo}SAX
% in 1997 and again in 1998.
% The solid circles in the IR are observations simultaneous with the
% 1998 X--ray data.
% The solid line is a one--zone homogeneous synchrotron self--Compton model,
% see Tagliaferri et al. 1999.}
% \end{figure}

\section{The re--united blazars}

Fossati et al. (1998) found that the SED of all blazars is related 
to their observed luminosity.
There is a rather well defined trend:
low luminosity objects are HBL--like, and furthermore their high energy
peak is in the GeV--TeV band.
As the bolometric luminosity increases, both peaks shift to lower frequencies,
and the high energy emission is increasingly more dominating the total output.
\footnote{A note of caution: the limited sensitivity of EGRET (onboard CGRO)
and ground based Cherenkov telescopes allows to detect sources which are
in high states. Therefore the trend of more high energy dominated spectra 
as the total power increases strictly refers to high states.} 
Ghisellini et al. (1998), fitted the SED of all blazars detected 
in the $\gamma$--ray band for which the distance and some spectral 
information of the high energy radiation were available.
They found a correlation between the energy $\gamma_{\rm peak}m_{\rm e}c^2$
of the electrons emitting at the peaks of the spectrum and the amount 
of energy density $U$ (both in radiation and in magnetic field), as measured 
in the comoving frame: $\gamma_{\rm peak}\propto U^{-0.6}$.
This indicates that, at $\gamma_{\rm peak}$, the radiative cooling rate 
$\dot\gamma(\gamma_{\rm peak})\propto \gamma_{\rm peak}^2 U \sim$const.
It also suggests that this may be due to a ``universal" acceleration
mechanism, which must be nearly independent of $\gamma$ and $U$:
in less powerful sources with weak magnetic field and weak lines the
radiative cooling is less severe and electrons can be accelerated up to
very high energies, producing a SED typical of a HBL.
The paucity of photons produced externally to the jet leaves synchrotron
self--Compton as the only channel to produce high energy radiation.
At the other extreme, in the most powerful sources with strong emission lines,
electrons cannot be accelerated to high energies because of severe cooling.
Their spectrum is therefore peaked in the far IR and in the MeV band.
In these sources the inverse Compton scattering off externally produced
photons is the dominant cooling mechanism, producing a dominant
$\gamma$--ray luminosity.

\subsection{Powers}

For the same sample of blazars fitted in Ghisellini et al. (1998)
we can estimate the powers radiated and transported 
by jets in the form of cold protons,
magnetic field and hot electrons and/or electron--positron pairs.
Since the model allows to determine the bulk Lorentz factor,
the dimension of the emitting region, the value of the magnetic
field and the particle density, we can then determine
\begin{equation}
L_{\rm p} \, = \, \pi R^2 \Gamma^2\beta c\, n_{\rm p}^\prime 
m_{\rm p} c^2;\quad
L_{\rm e}\,  = \, \pi R^2 \Gamma^2\beta c\, n_{\rm e}^\prime 
\langle \gamma \rangle m_{\rm e} c^2 ;\quad
L_{\rm B}\, = \, \pi R^2 \Gamma^2\beta c \, {B^2 \over 8\pi} 
\end{equation}
where $n_{\rm p}^\prime$ and $n_{\rm e}^\prime$ are the comoving proton and 
lepton densities,
respectively, $R$ is the cross section radius of the jet, and
$\langle \gamma \rangle m_{\rm e} c^2$ is the average lepton energy.
These powers can be compared with the radiated one estimated in the
same frame (in which the emitting blob is seen moving). 
The power radiated {\it in the entire solid angle} is thus 
$L_{\rm r}=L^\prime_{\rm r} \Gamma^2$ (the same holds for the power 
$L_{\rm syn}$ emitted by the synchrotron process).
All these quantities are plotted in Fig. 2
(Celotti \& Ghisellini 2000, in prep.).
In this figure hatched areas correspond to BL Lac objects.
Several facts are to be noted:
\begin{itemize}
\item If the jet is made by a pure electron--positron plasma,
then the associated kinetic power is $L_{\rm e}$. 
However, we note that $L_{\rm e} \ll L_{\rm r}$
posing a serious energy budget problem.
\item If there is a proton for each electron, the bulk kinetic
power $L_{\rm p}\sim 10  L_{\rm r}$.
This corresponds to an efficiency of $\sim 10\%$ in converting
bulk into random energy.
The remaining 90\% is therefore available to power the radio lobes,
as required.
\item The power in the Poynting flux, $L_{\rm B}$,
is of the same order of $L_{\rm e}$, indicating that the magnetic 
field is close to equipartition with the electron energy density.
This suggests that, on these scales, the magnetic field is not a prime 
energy carrier, but is a sub--product of the process transforming
bulk into random energy.
\end{itemize}

\begin{figure}
\vskip -1 true cm
\centerline{\psfig{file=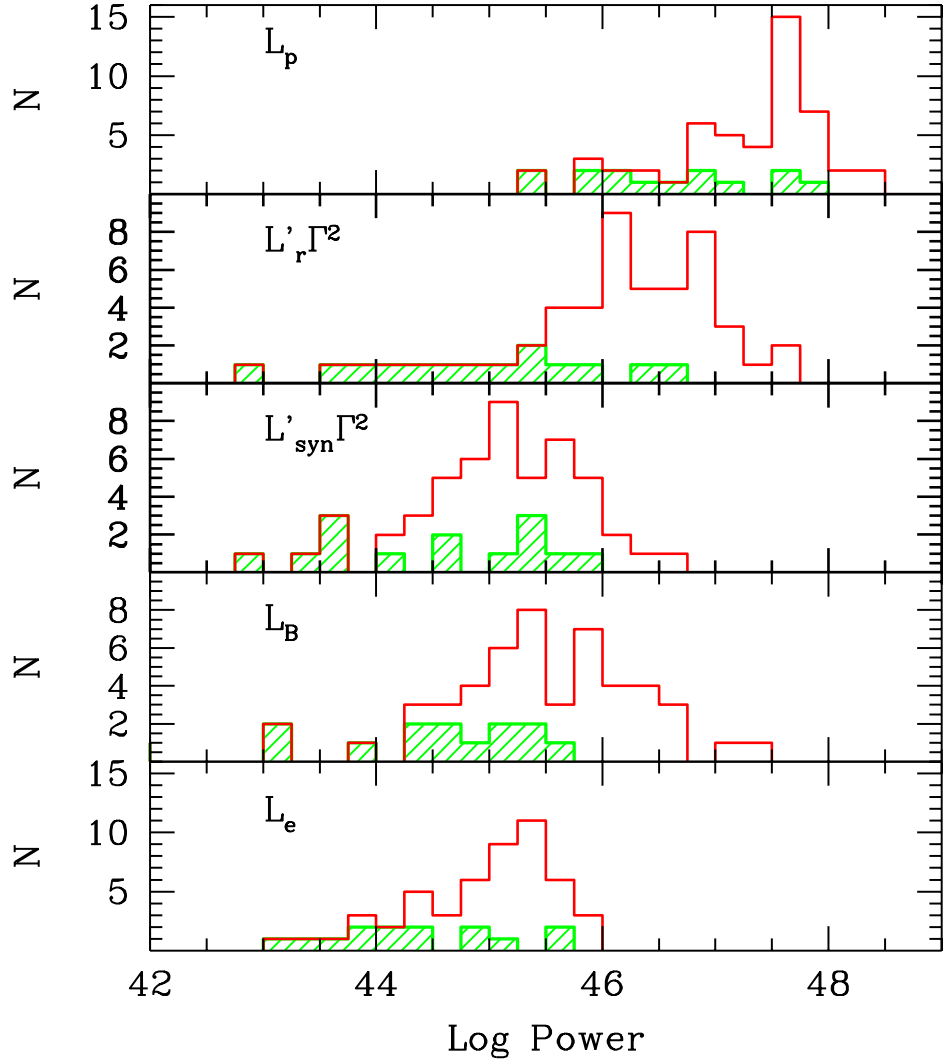, width=15cm}} % , height=9.0cm}}
\caption[]{Histograms of the powers carried by the jet in protons, 
total radiation, synchrotron radiation, magnetic field and 
relativistic electrons, from top to bottom.
Hatched areas correspond to BL Lac objects.
The electron distribution was assumed to extend down
to $\gamma_{\rm min}\sim 1$. From Celotti \& Ghisellini (2000, in prep.)}
\end{figure}

\begin{figure}
\vskip -2. true cm
\centerline{\psfig{file=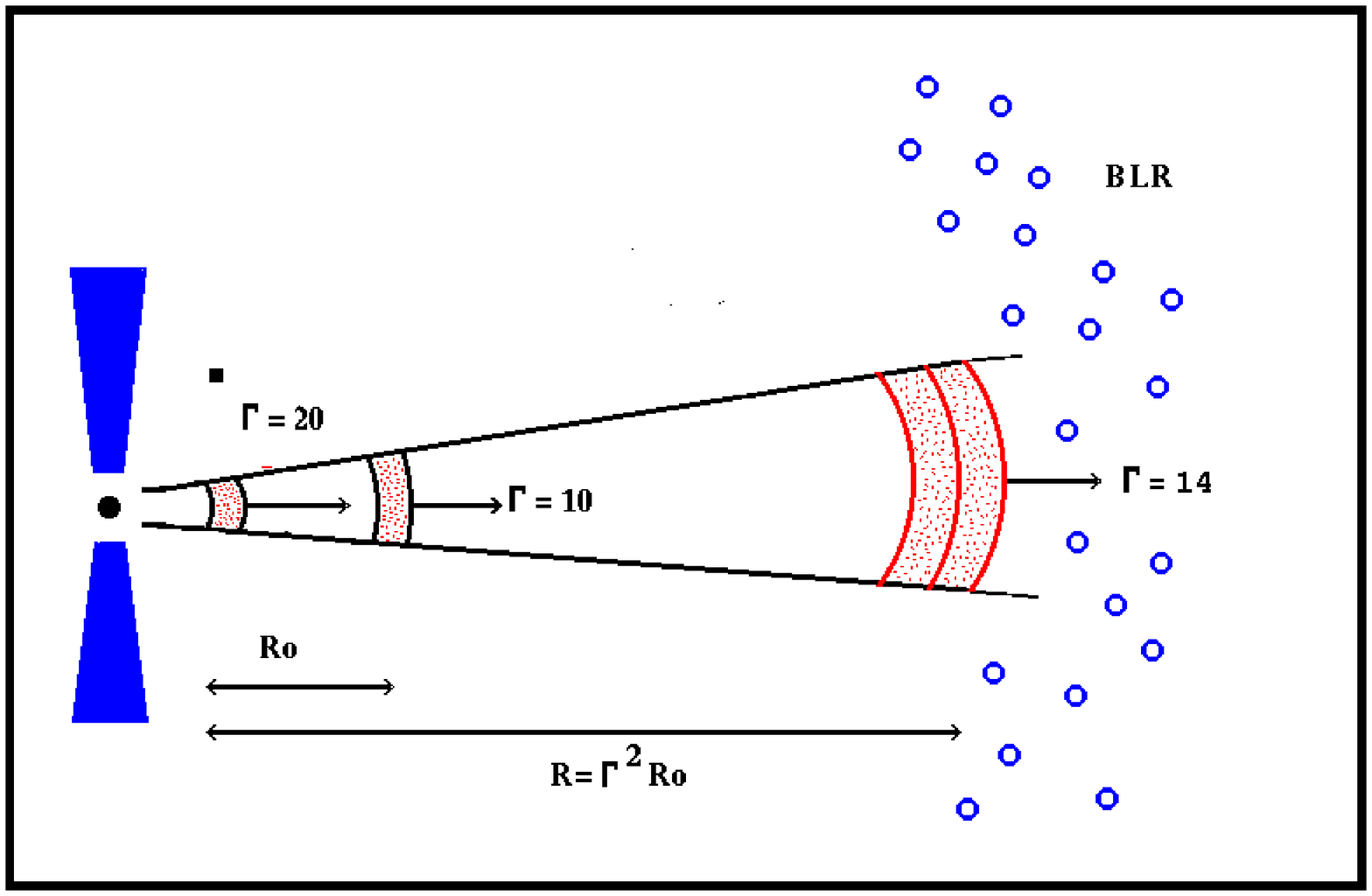, width=13cm}}
\vskip -7 true cm
\caption[]{Cartoon illustrating the internal shock scenario.
The intermittent activity of the central engine produces two shells,
initially separated by $R_0$. 
The faster one will catch up the slower one at $R\sim \Gamma^2 R_0$.}
\end{figure}

\section{Internal shocks}

The central engine may well inject energy into the jet in a discontinuous
way, with individual shells or blobs having different masses, bulk Lorentz
factors and energies.
If this occurs there will be collisions between shells, with a faster
shell catching up a slower one.
This idea has become the leading model to explain the emission of 
gamma--ray bursts, but it was born in the AGN field, due to Rees (1978)
(see also Sikora 1994).

\begin{itemize}
\item {\bf Location --- }
The $\gamma$--ray emission of blazars and its rapid variability
imply that there must be a preferred location where dissipation
of the bulk motion energy occurs. 
If it were at the base of the jet, and hence close to the accretion disk,
the produced $\gamma$--rays would be inevitably absorbed by 
photon--photon collisions, with associated copious pair production,
reprocessing the original power from the $\gamma$--ray to the X--ray
part of the spectrum (contrary to observations).
If it were far away, in a large region of the jet, it becomes difficult to 
explain the observed fast variability,
even accounting for the time--shortening due to the Doppler effect.
The region where the radiation is produced is then most likely located
at a few hundreds of Schwarzchild radii ($\sim 10^{17}$ cm) 
from the base of the jet, within the broad line region
(see Ghisellini \& Madau 1996 for more details).
The extra seed photons provided by emission lines enhance the efficiency 
of the Compton process responsible for the $\gamma$--ray emission. 
This is indeed the typical distance at which two shells, initially 
separated by $R_0\sim 10^{15}$ cm (comparable to a few Schwarzschild radii) 
and moving with $\Gamma \sim 10$ and $\Gamma\sim 20$ would collide.

\item {\bf Variability timescales ---}
In fact if the initial separation of the two shells is $R_0$ and
if they have Lorentz factors $\Gamma_1$, $\Gamma_2$, 
they will collide at
\begin{equation}
R\, = {2 \Gamma_1^2 \over 1-(\Gamma_1/\Gamma_2)^2} \,  R_0 
\end{equation}
If the shell widths are of the same order of their initial separation
the time needed to cross each other is of the order of $R/c$.
The observer at a viewing angle $\theta\sim 1/\Gamma$ will see this time
Doppler contracted by the factor $(1-\beta\cos\theta)\sim \Gamma^{-2}$.
The typical variability timescale is therefore of the same order
of the initial shell separation.
If the mechanism powering GRB and blazar emission is the same,
we should expect a similar light curve from both systems, 
but with times appropriately scaled by the different $R_0$, i.e. 
the different masses of the involved black holes.

\item {\bf Efficiencies ---}
As most of the power transported by the jet must reach the radio lobes, 
only a small fraction can be radiatively dissipated.
The efficiency $\eta$ of two blobs/shells for converting ordered into
random energy depends on their masses $m_1$, $m_2$ and 
bulk Lorentz factors $\Gamma_1$, $\Gamma_2$, as
\begin{equation}
\eta \,=\, 1-\Gamma_f\, { m_1+m_2\over \Gamma_1 m_1 +\Gamma_2m_2}
\end{equation}
where $\Gamma_f=(1-\beta_f^2)^{-1/2}$ is the bulk Lorentz factor 
after the interaction and is given by 
(see e.g. Lazzati, Ghisellini \& Celotti 1999)
\begin{equation}
\beta_f = {\beta_1\Gamma_1m_1+ \beta_2\Gamma_2m_2 \over
\Gamma_1m_1+ \Gamma_2m_2}
\end{equation}
The above relations imply, for shells of equal masses and 
$\Gamma_2=2\Gamma_1=20$, $\Gamma_f=14.15$ and $\eta=5.7\%$.

Efficiencies $\eta$ around 5--10\% are just what needed
for blazar jets.

\item {\bf Peak energies? ---}
In the rest frame of the fast shell, the bulk kinetic energy
of each proton of the slower shell is $\sim (\Gamma^\prime-1)m_pc^2$, 
where $\Gamma^\prime\sim 2$.  
This is what can be transformed into random energy.
Assume now that the electrons share this available energy
(through an unspecified acceleration mechanism).
In the comoving frame, the acceleration rate can be written as 
$\dot E_{heat} \sim (\Gamma^\prime-1)m_p c^2 /t^\prime_{heat}$.
The typical heating timescale may correspond to the time needed for the 
two shells to cross, i.e. 
$t^\prime_{heat}\sim \Delta R^\prime/c\sim R/(c\Gamma)$,
where $\Delta R^\prime$ is the shell width (measured in the same frame).
The heating and the radiative cooling rates will balance for some
value of the random electron Lorentz factor $\gamma_{peak}$:
\begin{equation}
\dot E_{heat}\, =\,  \dot E_{cool}\, \to \,
{\Gamma m_p c^3 \over R} \, =\, {4\over 3} \sigma_T c U\gamma^2_{peak}
\, \to \,
\gamma_{peak} \, =\, \left({ 3\Gamma m_p c^2 \over 4 \sigma_T R U}\right)^{1/2}
\end{equation}
The agreement  of the above simple relation with what
can be derived from model fitting the SED of blazars is surprisingly good
(see Ghisellini 2000).

\item {\bf Radio flares ---}
Collisions between shells may (and should) happen in a hierarchical way.
As an illustrative example, assume that one pair of shells
after the collision moves with a final Lorentz factor $\Gamma_1 =14$
(this number corresponds to $\Gamma=10$ and 20 for the two shells
before the interaction).
The collision produces a flare --say-- in the optical and $\gamma$--ray bands.
After some observed time $\Delta t$ two other shells collide and another
flare is produced.
Assume that the final Lorentz factor is now $\Gamma_2=17$ (corresponding to 
an initial $\Gamma =10$ and 30 before collision).
Since the second pair is faster, it will catch up the first
one after a distance (from eq. 5) $R \sim  1200 c \Delta t$.
A time separation of $\Delta t \sim$ a day between the two flares 
then corresponds to $R\sim$ 1 pc, i.e. the region of the radio emission
of the core.
Due again to Doppler contraction, this radio flare will be observed
ony a few days after the second optical flare.
Since the ratio $\Gamma_2/\Gamma_1$ is small, the efficiency is also small
(at least a factor 10 smaller than the firsts shocks).
There is then the intriguing possibility of explaining the birth of radio
blobs after intense activity (i.e. more than one flare) of the
higher energy flux.
Radio light--curves should have {\it some} memory of what has happened
days--weeks earlier at higher frequencies.

\end{itemize}

\section{Conclusions}

Here I will dare to assemble different pieces of information
gathered in recent years in a coherent, albeit still preliminary, picture.

There is a link between the extraction of gravitational energy
in an accretion disk and the formation and acceleration of jets,
since both have the same power.
Objects of low luminosity accretion disks also lack strong emission lines,
suggesting that it is the paucity of ionizing photons, not of gas, the reason
for the lack of strong lines in BL Lacs.
Correspondingly, this implies that, if FR I are the parents of BL Lacs,
they also have intrinsically weak line emission 
(i.e. no need for an obscuring torus).
Despite the fact that the jet power in blazars spans at least four orders
of magnitude, the average bulk Lorentz factor is almost the same,
suggesting a link between the power and the mass outflowing rate:
their ratio is constant.
In the region where most of the radiation is produced, the jet is heavy,
in the sense that protons carry most of the bulk kinetic energy.
There the jet dissipates $\sim 10\%$ of its power and produces
beamed radiation.
The power dissipated at larger distances is much less, and therefore
the jet can transport $\sim 90\%$ of its original power to 
the radio extended regions.
One way to achieve this is through internal shocks, which can explain
why the major dissipation occurs at a few hundreds Schwarzchild radii,
why the efficiency is of the order of 10\%, and give clues
on the observed variability timescales and even on why electrons
are accelerated at a preferred energy.
The spectral energy distribution of blazars depends on where 
shell--shell collisions take place, and on the amount of seed
photons present there.
Even in a single source it is possible that the separation 
of two consecutive shells is sometimes large, resulting in a collision
occuring outside the broad line region.
In this case the corresponding spectrum should be produced
by the synchrotron self--Compton process only, without the contribution
of external photons: we then expect a simultaneous 
optical--$\gamma$--ray flare of roughly equal powers (but with
the self--Compton flux varying quadratically, see Ghisellini \& Maraschi 1966).
This is what should always happen in lineless BL Lac objects.
On the other hand, if the initial separation of the two shells
is small (or the $\Gamma$--factor of the slower one is small),
the collision takes place close to the disk.
X--rays produced by the disk would then absorb all the produced 
$\gamma$--rays and a pair cascade would develop, reprocessing the power 
originally in the $\gamma$--ray band mainly into the X--ray band.
We should therefore see an X--ray flare without accompanying 
emission above $\Gamma m_{\rm e} c^2$.

Pairs of shells which have already collided can interact again
between themselves, at distances appropriate for the radio
emission.
This offers the interesting possibility to explain why the radio 
luminosity is related with the $\gamma$--ray one, and why radio 
flares are associated with flares at higher frequencies.
Work is in progress in order to quantitatively test
this idea against observations.

\begin{acknowledgements}
I thank Annalisa Celotti for very insightful discussions.
\end{acknowledgements}

\end{document}